\documentclass[prd,twocolumn,preprintnumbers,superscriptaddress,nofootinbib,a4paper]{revtex4}

\usepackage{amstext,amssymb}
\usepackage{amsmath}
\usepackage{graphicx}
\usepackage[hyperfootnotes=false]{hyperref}
\usepackage{xspace}
\usepackage{color}
\usepackage{units}
\usepackage{slashed} 

\begin{document}
\title{Neutrinoless Double Beta Decay in Left-Right Symmetry with Universal Seesaw}
\author{Frank F. \surname{Deppisch}}
\email{f.deppisch@ucl.ac.uk}
\affiliation{Department of Physics and Astronomy, University College London, London WC1E 6BT, United Kingdom}

\author{Chandan \surname{Hati}}
\email{chandan@prl.res.in}
\affiliation{Physical Research Laboratory, Navrangpura, Ahmedabad 380009, India}
\affiliation{Indian Institute of Technology Gandhinagar, Chandkheda, Ahmedabad 382424, India}

\author{Sudhanwa \surname{Patra}}
\email{sudha.astro@gmail.com}
\affiliation{Center of Excellence in Theoretical and Mathematical Sciences, 
Siksha 'O' Anusandhan University, Bhubaneswar 751030, India}

\author{Prativa \surname{Pritimita}} 
\email{pratibha.pritimita@gmail.com}
\affiliation{Center of Excellence in Theoretical and Mathematical Sciences, 
Siksha 'O' Anusandhan University, Bhubaneswar 751030, India}

\author{Utpal \surname{Sarkar}}
\email{utpal@phy.iitkgp.ernet.in}
\affiliation{\hspace*{-0.0cm}Department of Physics, Indian Institute of Technology Kharagpur, Kharagpur 721302, India}

\begin{abstract}
\noindent 
We discuss a class of left-right symmetric theories with a universal seesaw mechanism for fermion masses and mixing and the implications for neutrinoless double beta ($0\nu\beta\beta$) decay where neutrino masses are governed by natural type-II seesaw dominance. The scalar sector consists of left- and right-handed Higgs doublets and triplets, while the conventional Higgs bidoublet is absent in this scenario. We use the Higgs doublets to implement the left-right and the electroweak symmetry breaking. On the other hand, the Higgs triplets with induced vacuum expectation values can give Majorana masses to light and heavy neutrinos and mediate $0\nu\beta\beta$ decay. In the absence of the Dirac mass terms for the neutrinos, this framework can naturally realize type-II seesaw dominance even if the right-handed neutrinos have masses of a few TeV. We study the implications of this framework in the context of $0\nu\beta\beta$ decay and gauge coupling unification. 
\end{abstract}

\maketitle

\noindent
\section{Introduction}
\label{sec1}
The neutrino oscillation experiments have established that the neutrinos have nonzero masses. However, the question regarding the fundamental nature of the neutrinos; whether they are Dirac~\cite{Dirac:1925} or Majorana~\cite{Majorana:1937vz}, is yet to be answered. From a theoretical point of view several frameworks predict Majorana neutrinos, while the experimental searches are still inconclusive in this regard. To this end the detection of neutrinoless double beta ($0\nu\beta\beta$) decay, which requires neutrinos to be Majorana particles regardless of the underlying mechanism, plays the crucial role in confirming the nature of the neutrinos. This rare process, the conversion of two neutrons into two protons, two electrons and nothing else, if observed, would pave a path towards the search for new physics beyond the Standard Model (SM). Currently, the KamLAND-Zen experiment \cite{KamLAND-Zen:2016pfg} (using $^{136}\mbox{Xe}$) claims the best limit on the half-life $T_{1/2}$ to be less than $1.07\times 10^{26}$~yr corresponding to an upper bound on effective Majorana mass $m_{\mathrm{eff}} \lesssim 0.06 - 0.17~{\rm eV}$, depending on the nuclear matrix element calculation used. 

Models naturally accommodating neutrino masses are the need of the time and the left-right symmetric model (LRSM)~\cite{Pati:1974yy, Mohapatra:1974gc, Senjanovic:1975rk, Senjanovic:1978ev,Mohapatra:1979ia, Mohapatra:1980yp} is one of the most popular candidates for this purpose. The right-handed neutrinos and mirror gauge bosons present make it quintessential to explain the $V-A$~nature of the weak interactions, tiny but nonzero masses of neutrinos~\cite{Minkowski:1977sc, GellMann:1980vs, Yanagida:1979as, Schechter:1980gr, Schechter:1981cv} and a lot more. Moreover, the spontaneous breaking of the left-right symmetry at TeV scale offers a plethora of possibilities in collider phenomenology. The LHC gives a lower bound on the right-handed charged gauge boson mass $M_{W_R}$ of $\mathcal{O}(3\text{ TeV})$~\cite{Khachatryan:2014dka}, while the Kaon $K_L-K_S$ mass difference results in a lower bound on $M_{W_R}$ of $2.5$~TeV \cite{Beall:1981ze, Zhang:2007da, Maiezza:2010ic, Bertolini:2014sua}\footnote{Lower values of $M_{W_R}$ in the range $1.9-2.5$~TeV are allowed within TeV scale LRSMs with spontaneous D-parity breaking~\cite{Deppisch:2014qpa, Deppisch:2014zta, Deppisch:2015cua}.}. Such a low scale $W_R$ gauge boson associated with right-handed charged currents can give rise to new contributions to $0\nu\beta\beta$ decay and can be accessible at the LHC.  

In this work we study a LRSM framework with vector-like fermions realizing a universal seesaw mechanism for fermion masses, except for the neutrinos. The scalar sector consists of left- and right-handed Higgs doublets and triplets, while the conventional Higgs bidoublet is absent in this scenario. We use the Higgs doublets to implement the left-right and the electroweak symmetry breaking. On the other hand the Higgs triplets with induced vacuum expectation values give Majorana masses to the light and heavy neutrinos in the absence of the Dirac mass terms. Consequently, one can naturally realize a type-II seesaw dominance in this framework even if the right-handed neutrinos have masses around a few TeV. We study its implications for the $0\nu\beta\beta$ decay as well as the possibility of gauge coupling unification in this framework.

The outline for the rest of this paper is as follows. In section~\ref{sec2} we start with a brief overview of the model, followed by discussions on the generation of fermion masses via universal seesaw in section~\ref{sec3}, neutrino masses via type-II seesaw dominance in section~\ref{sec4} and the gauge bosons in section~\ref{sec5}. In section~\ref{unifn}, we discuss the model in the context of gauge coupling unification. In section~\ref{secbeta} we discuss the implications for $0\nu\beta\beta$ decay. Finally, in section \ref{conclusion} we summarize our results and conclude.

\section{Left-right symmetry with vector-like fermions}
\label{sec2}

The gauge group for LRSMs is $SU(2)_L \times SU(2)_R \times U(1)_{B-L} \times SU(3)_C$. 
The usual fermion content of the model is
\begin{gather}
	Q_{L}=\begin{pmatrix}u_{L}\\
	d_{L}\end{pmatrix}\equiv[2,1,{\frac{1}{3}},3], \quad Q_{R}=\begin{pmatrix}u_{R}\\
	d_{R}\end{pmatrix}\equiv[1,2,{\frac{1}{3}},3]\,,\nonumber \\
	\ell_{L}=\begin{pmatrix}\nu_{L}\\
	e_{L}\end{pmatrix}\equiv[2,1,-1,1], \quad 
	\ell_{R}=\begin{pmatrix}\nu_{R}\\
	e_{R}\end{pmatrix}\equiv[1,2,-1,1] \,, 
\end{gather}
where the numbers in brackets correspond to the transformations under $SU(2)_L \times SU(2)_R \times U(1)_{B-L} \times SU(3)_C$. We also consider additional vector-like quarks and charged leptons~\cite{Dev:2015vjd, Deppisch:2016scs, Patra:2012ur},
\begin{gather}
	U_{L,R}\equiv[1,1,4/3,3]\,,\quad D_{L,R}\equiv[1,1,-2/3,3]\, , \nonumber \\
	E_{L,R}\equiv[1,1,-2,3].
\end{gather}

We implement a scalar sector consisting of $SU(2)_{L,R}$ doublets and triplets, however the conventional scalar bidoublet is absent. We use the Higgs doublets to implement the left-right and the electroweak symmetry breaking: $H_R \equiv (h_R^0, h^-_R)^T \equiv [1,2,-1,1]$ breaks the left-right symmetry, while $H_L\equiv (h_L^0, h^-_L)^T \equiv [2,1,-1,1]$ breaks the electroweak symmetry once they acquire vacuum expectation values (VEVs),
\begin{align}
	\langle H_R \rangle = \begin{pmatrix} \frac{v_R}{\sqrt{2}} \\ 0 \end{pmatrix}, \quad 
	\langle H_L \rangle = \begin{pmatrix} \frac{v_L}{\sqrt{2}} \\ 0 \end{pmatrix}. 
\end{align}
Note that the present framework requires only doublet Higgs fields for spontaneous symmetry breaking. However, in the absence of a Higgs bidoublet, we use the vector-like new fermions to generate correct charged fermion masses through a universal seesaw mechanism. For the neutrinos we note that in the absence of a scalar bidoublet there is no Dirac mass term for light neutrinos and without scalar triplets no Majorana masses are generated either. To remedy this fact we introduce additional scalar triplets $\Delta_{L}$ and $\Delta_{R}$,
\begin{align}
	\Delta_{L,R} &= \begin{pmatrix} \delta_{L,R}^+/\sqrt{2} & \delta_{L,R}^{++} \\ \delta_{L,R}^0 & -\delta_{L,R}^+/\sqrt{2} \end{pmatrix}\,,
\end{align}
which transform as $\Delta_L \equiv [3,1,-2,1]$ and $\Delta_R \equiv [1,3,-2,1]$, respectively. They generate Majorana masses for the light and heavy neutrinos although they are not essential in spontaneous symmetry breaking here. In the presence of the Higgs triplets, the manifestly Left-Right symmetric scalar potential has the form
\begin{align}
\mathcal{L}&  = 
   (D_\mu H_L)^\dagger D^\mu H_L 
 + (D_\mu H_R)^\dagger D^\mu H_R \nonumber\\
&+ (D_\mu \Delta_L)^\dagger D^\mu \Delta_L 
 + (D_\mu \Delta_R)^\dagger D^\mu \Delta_R \nonumber\\
&- \mu^2_H (|H_L|^2 + |H_R|^2) 
 - \lambda \left( |H_L|^4 +  |H_R|^4 \right) \nonumber\\
&+ \mu^2_\Delta (|\Delta_L|^2 + |\Delta_R|^2) 
 - \lambda^\prime \left( |\Delta_L|^4 + |\Delta_R|^4 \right) \nonumber \\
&-\beta_1 |H_L|^2 |H_R|^2- \rho_1 (|\Delta_L|^2 |H_R|^2+|\Delta_R|^2 |H_L|^2) \nonumber \\
&- \rho_{2} \left( |\Delta_L|^2 |H_L|^2+|\Delta_R|^2 |H_R|^2 \right) \nonumber \\
& -\rho_{3}\left( H_L^\dagger \Delta_L^\dagger \Delta_L H_L +H_R^\dagger \Delta_R^\dagger \Delta_R H_R\right) \nonumber \\
&- \mu_{} \left( H^T_L i\sigma_2 \Delta_L H_L+H^T_R i\sigma_2 \Delta_R H_R \right) +\mbox{h.c.}\,.
\label{eq:slepton_lagrangian}
\end{align}
After the Higgs doublets $H_R$ and $H_L$ acquire their VEVs, the Higgs triplets get induced VEVs,
\begin{align}
	\langle \Delta_L \rangle \equiv u_L = \frac{\mu v_L^2}{M^2_{\delta_L^0}}\,, \quad
	\langle \Delta_R \rangle \equiv u_R = \frac{\mu v_R^2}{M^2_{\delta_R^0}}\,.
\end{align}
%

\section{Fermion Masses via universal seesaw}
\label{sec3}
As discussed earlier, in this scheme normal Dirac mass terms for the SM fermions are not allowed due to the absence of a bidoublet Higgs scalar. However, in the presence of vector-like copies of quark and charged lepton gauge isosinglets, the charged fermion mass matrices can assume a seesaw structure. The Yukawa interaction Lagrangian in this model is given by
\begin{align}
\mathcal{L} = 
	&- Y_U^L \tilde{H}_L \overline{q}_L U_R 
	 + Y_U^R \tilde{H}_R \overline{q}_R U_L +Y_D^L H_L \overline{q}_L D_R \nonumber\\
	&+ Y_D^R H_R \overline{q}_R D_L + Y_E^L H_L\overline{\ell}_L E_R 
	 + Y_E^R H_R\overline{\ell}_R E_L \nonumber\\
	&+\frac{1}{2}f \left(\overline{\ell^c_L} i\tau_2 \Delta_L \ell_L 
	 + \overline{\ell^c_R} i\tau_2 \Delta_R \ell_R \right) \nonumber \\
	&- M_U \overline{U} U - M_D \overline{D} D - M_E \overline{E} E + \text{h.c.},
\label{2.1}
\end{align}
where we suppress the flavor and color indices on the fields and couplings. $\tilde{H}_{L,R}$ denotes $\tau_2 H_{L,R}^\ast$, where $\tau_2$ is the usual second Pauli matrix. Note that there is an ambiguity regarding the breaking of parity, which can either be broken spontaneously with the left-right symmetry at around the TeV scale or at a much higher scale independent of the left-right symmetry breaking. In the latter case, the Yukawa couplings corresponding to the right-type and left-type Yukawa terms can be different because of the renormalization group running below the parity breaking scale, $Y_X^R \neq Y_X^L$. Thus, while writing the Yukawa terms above we distinguish the left- and right-handed couplings explicitly with the subscripts $L$ and $R$.

\begin{figure*}[t!]
\includegraphics[width=0.99\textwidth]{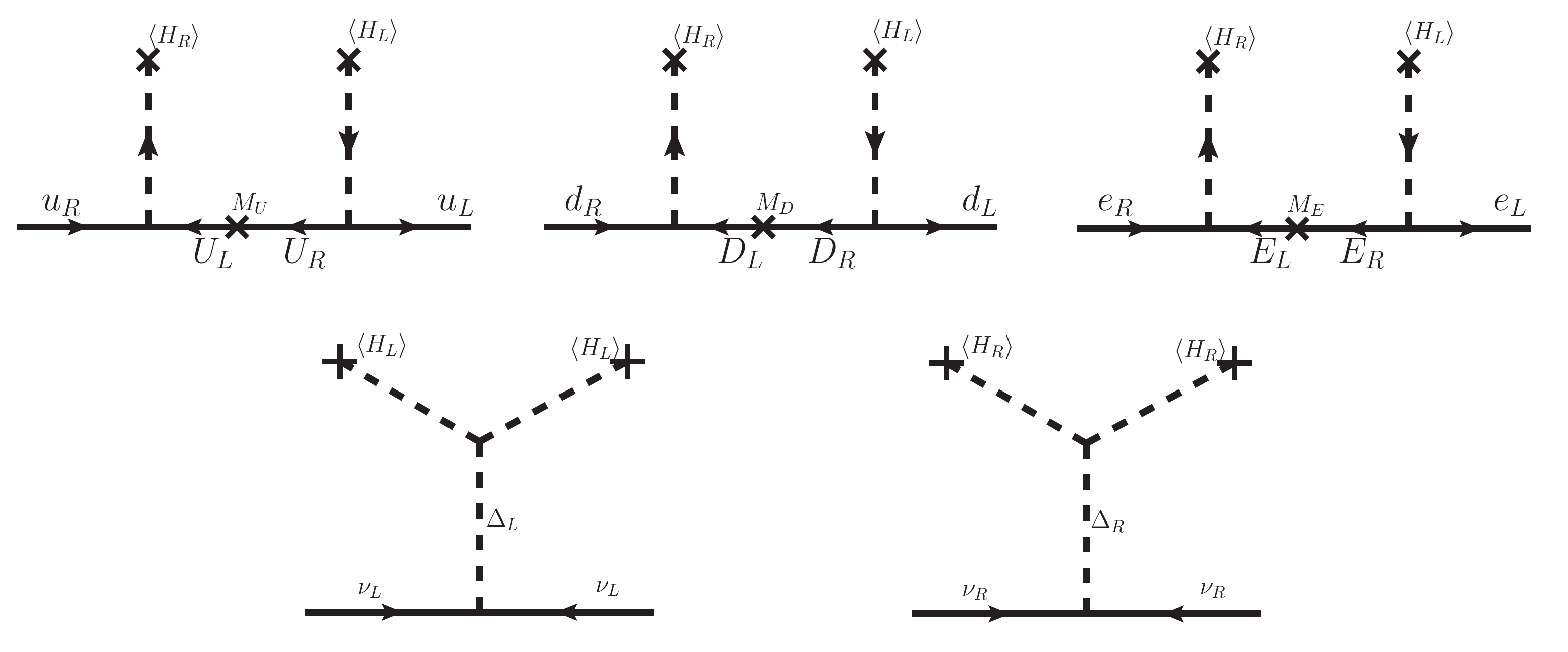}
\caption{Generation of fermion masses through universal seesaw and induced triplet VEVs.}
\label{feyn:seesaw}
\end{figure*}
After spontaneous symmetry breaking we can write the mass matrices for the charged fermions as \cite{Deppisch:2016scs}
\begin{gather}
\label{2.3}
	M_{uU}    = \begin{pmatrix} 0 & Y_U^L v_L \\ Y_U^R v_R & M_U \end{pmatrix}, \,
	M_{dD}    = \begin{pmatrix} 0 & Y_D^L v_L \\ Y_D^R v_R & M_D \end{pmatrix}, \nonumber\\
	M_{e E}   = \begin{pmatrix} 0 & Y_E^L v_L \\ Y_E^R v_R & M_E \end{pmatrix}.
\end{gather}
The corresponding generation of fermion masses is diagrammatically depicted in Fig.~\ref{feyn:seesaw}.

Assuming all parameters to be real one can obtain the mass eigenstates by rotating the mass matrices via left and right orthogonal transformations ${\cal{O}}^{L,R}$. For example, up to leading order in $Y^L_U v_L$, the SM and heavy vector partner up-quark masses are
\begin{align}
\label{2.4.0}
	m_u \approx Y^L_U Y^R_U \frac{v_L v_R}{\hat{M}_U}, \quad
	\hat{M}_U \approx \sqrt{M_U^2 + (Y^R_U v_R)^2},
\end{align}
and the mixing angles $\theta^{L,R}_U$ in ${\cal{O}}^{L,R}$ are determined as
\begin{align}
\label{2.4.1}
	\tan(2\theta^{L,R}_U) \approx 2 Y^{L,R}_U \frac{v_{L,R} M_U}{M_U^2 \pm (Y^R_U v_R)^2}.
\end{align}
The other fermion masses and mixing are obtained in an  analogous manner. Note that here we have neglected the flavor structure of the Yukawa couplings $Y^{L,R}_X$ which will determine the observed quark and charged lepton mixings. The hierarchy of SM fermion masses can be explained by assuming either a hierarchical structure of the Yukawa couplings or a hierarchical structure of the vector-like fermion masses.

\section{Neutrino Masses and type II seesaw dominance}
\label{sec4}

In the model under consideration there is no tree level Dirac mass term for the neutrinos due to the absence of a Higgs bidoublet. The scalar triplets acquire induced VEVs $\langle \Delta_L \rangle =u_L$ and $\langle \Delta_R \rangle =u_R$ giving the neutral lepton mass matrix in the basis $(\nu_L, \nu_R)$ given by
\begin{align}
	M_\nu= 
	\left(\begin{array}{cc}
		f u_L  & 0  \\
   	    0 & f u_R
\end{array} \right) \,.
\label{eqn:numatrix}       
\end{align}
Thus the light and heavy neutrino masses are simply $m_\nu = f u_L \propto M_N = f u_R$.  A Dirac mass term is generated at the two-loop level via the one-loop $W$ boson mixing $\theta_W$ (see the next section) and the exchange of a charged lepton. It is of the order $m_D \lesssim g_L^4/(16\pi^2)^2 m_\tau m_b m_t / M_{W_R}^2 \approx 0.1$~eV for $M_{W_R} \approx 5$~TeV. This is intriguingly of the order of the observed neutrino masses; as long as the right-handed neutrinos are much heavier than the left-handed neutrinos, the type-II seesaw dominance is preserved and the induced mixing $m_D/M_N$ is negligible. The mixing between charged gauge bosons $\theta_W \approx g_L^2/(16\pi^2) m_b m_t/M^2_{W_R}$ is generated through the exchange of bottom and top quarks, and their vector-like partners. This yields a very small mixing of the order  $\theta_W \approx 10^{-7}$ for TeV scale $W_R$ bosons.

An interesting situation arises if we assume $u_L, u_R \ll v_L \ll v_R$. Then one can allow right-handed Majorana neutrinos with masses below GeV which can play an important role in $0\nu\beta\beta$ decay. Incorporating three fermion generations leads to the mixing matrices for the left- and right-handed matrices which we take to be equal
\begin{align}
  V_N = V_\nu \equiv U \,,
\label{eq:equality-VLR}
\end{align}
where $U$ is the phenomenological PMSN mixing matrix. Thus the unmeasured mixing matrix for the right-handed neutrinos is fully determined by the left-handed counterpart. The present framework gives a natural realization of type-II seesaw providing a direct relation between light and heavy neutrinos, $M_i \propto m_i$, i.e. the heavy neutrino masses $M_i$ can be expressed in terms of the light neutrino masses $m_i$ as $M_i = m_i (M_3/m_3)$, for a normal and $M_i = m_i (M_2/m_2)$ for a inverse hierarchy of light and heavy neutrino masses.

\section{Gauge bosons}
\label{sec5}
As discussed in the previous section, we consider a scenario where the VEVs of the Higgs doublets are much larger than the VEVs of the Higgs triplets i.e, $u_L \ll v_L, u_R \ll v_R$. Thus, the masses for the gauge bosons get small corrections from scalar triplets. The mass matrix for charged gauge bosons is given by
\begin{align}
  	M^2_W
	=
  	\frac{1}{4} 
  	\left(\begin{matrix}
  		g^2_L\left(v^2_L + 2 u^2_L\right)  & 0 \\
  		0 & g^2_R \left(v^2_R + 2 u^2_R \right)
	\end{matrix}\right) \,,
\end{align}
with the gauge couplings $g_L$ and $g_R$ associated with $SU(2)_L$ and $SU(2)_R$, respectively. The tree-level mixing between charged gauge bosons $W_L$ and $W_R$ is zero due to absence of a scalar bidoublet. The physical masses for charged gauge bosons are thus easily found,
\begin{align}
  M^2_{W_1} \approx \frac{g^2_L}{4} v^2_L \,,\quad
  M^2_{W_2} \approx \frac{g^2_R}{4} (v^2_R + 2 u^2_R) \,,
\end{align}
where we neglect $u_L$. At the one-loop level, a mixing of the order $\theta_W \approx g_L^2/(16\pi^2) m_b m_t/M_{W_R}^2$ is generated through the exchange of bottom and top quarks, and their vector-like partners. This yields a very small mixing of the order  $\theta_W \approx 10^{-7}$ for $M_{W_R} \approx 5$~TeV.

On the other hand, the neutral gauge boson mass matrix is given by
\begin{align}
	M^2_Z
	=
  	\frac{1}{4}
	\left(\begin{matrix}
  		    g^2_L v_L^2 
		&   0
		& - g_L g_{BL}\mu_L^2 \\
 		    0
		&   g^2_R \mu_R^2 
		& - g_{BL} g_R \mu_R^2 \\
  		  - g_{BL} g_L \mu_L^2 
		& - g_{BL} g_R \mu_R^2
		&   g^2_R \mu_L^2 + g^2_{BL} \mu_R^2
	\end{matrix}\right) \,,
\end{align}
with $\mu_{L,R}^2 = v_{L,R}^2 +4 u_{L,R}^2$ and the gauge coupling $g_{BL}$ associated with $U(1)_{B-L}$. The diagonalization gives mass eigenvalues for the neutral gauge bosons,
\begin{align}
  M^2_{Z_1} \approx \frac {g^2_L}{4 c^2_W} v^2_L\,,\quad
  M^2_{Z_2} \approx \frac {g^2_{BL} + g^2_R} 4 (v^2_R + 4 u^2_R) \,.
\end{align}
%

\section{Gauge coupling unification}{\label{unifn}} 
In this section we explore whether the LRSM framework under consideration can be embedded in a non-SUSY $SO(10)$ GUT theory unifying the gauge couplings. The possibility of achieving a gauge coupling unification in the presence of vector-like particles in a LRSM framework was studied in Ref.~\cite{Deppisch:2016scs}. However, here the Higgs sector and the vector-like fermion sector is different and thus it is worthwhile to study the gauge coupling unification in this framework. We consider a symmetry breaking pattern of $SO(10)$ such that it has the LRSM gauge group as its only intermediate symmetry breaking step as follows
\begin{align}
	SO(10) \mathop{\longrightarrow}^{\langle \Sigma \rangle} \mathcal{G}_{2213P}  
		   \mathop{\longrightarrow}^{\langle H_R    \rangle} \mathcal{G}_\text{SM} 
		   \mathop{\longrightarrow}^{\langle H_{L}  \rangle} 
		          \mathcal{G}_{U(1)_Q \times SU(3)}. 
\end{align}
The $SO(10)$ group breaks down to LRSM group $\mathcal{G}_{2213P} \equiv SU(2)_L \times SU(2)_R \times U(1)_{B-L} \times SU(3)$ via a non-zero VEV of $\Sigma \subset 210_H$~\cite{Chang:1983fu,Chang:1984uy,Chang:1984qr}. The particle content of the framework has been discussed in Sec-2. For successful gauge coupling unification we need to additionally introduce a scalar bitriplet $\eta \equiv (3,3,0,1)$ between the left-right symmetry breaking scale and the unification scale.

\begin{figure}[t!]
\includegraphics[width=0.95\columnwidth]{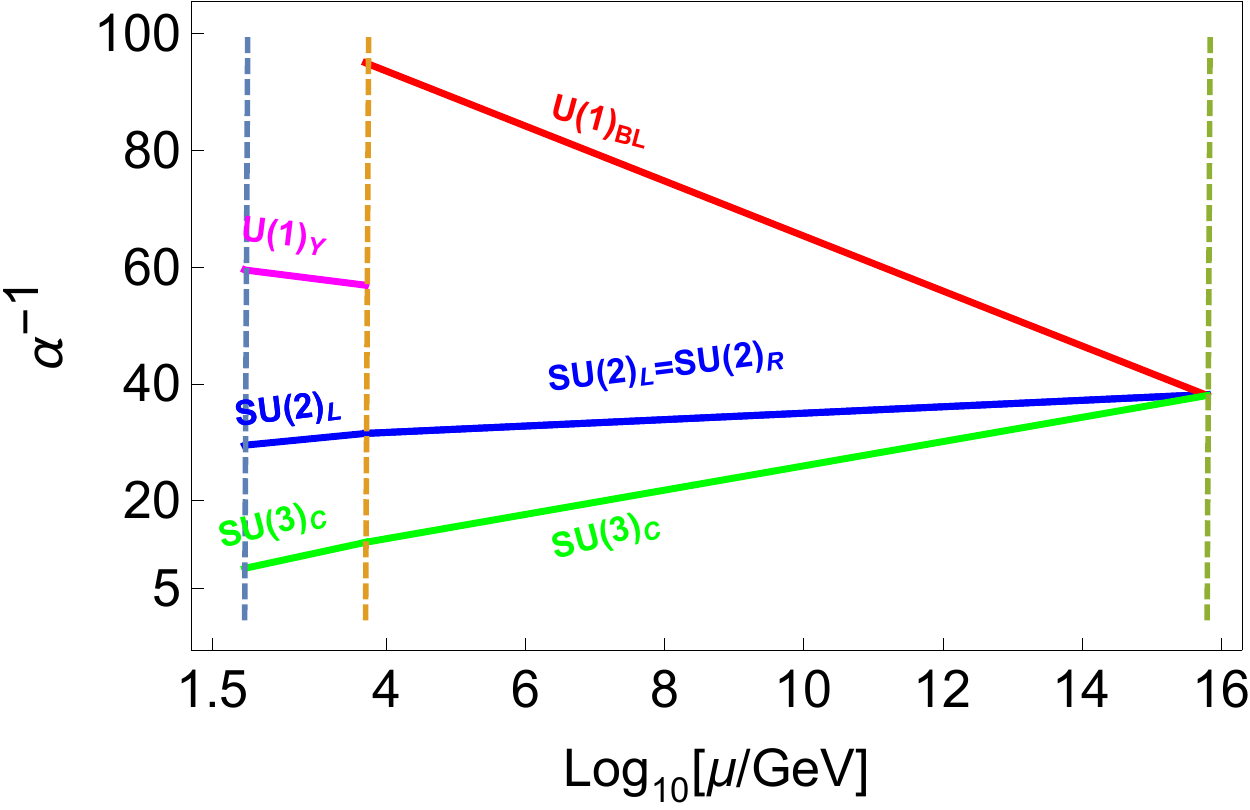}
\caption{Gauge coupling running and unification in the model considered with $M_\text{GUT} = 10^{15.75}$~GeV and the left-right symmetry breaking scale $M_\text{R} = 8$~TeV.}
\label{fig:unifn}
\end{figure}
The one-loop renormalization group equations (RGEs) for gauge couplings $g_i$ is given by
\begin{equation}
\mu\,\frac{\partial g_{i}}{\partial \mu}=\frac{b_i}{16 \pi^2} g^{3}_{i},
\end{equation}
where the one-loop beta-coefficients $b_i$ are given by
\begin{align}
	b_i &= - \frac{11}{3} \mathcal{C}_{2}(G) 
			+ \frac{2}{3} \,\sum_{R_f} T(R_f) \prod_{j \neq i} d_j(R_f) \nonumber \\
      &\qquad\qquad\qquad+ \frac{1}{3} \sum_{R_s} T(R_s) \prod_{j \neq i} d_j(R_s)\,.
\label{oneloop_bi}
\end{align}
Here, $\mathcal{C}_2(G)$ is the quadratic Casimir invariant, $T(R_f)$ and $T(R_s)$ are the traces of the irreducible representation $R_{f,s}$ for a given fermion and scalar, respectively. $d(R_{f,s})$ is the dimension of a given fermion/scalar representation $R_{f,s}$ under all $SU(N)$ gauge groups except the $i$-th~gauge group. For a real Higgs representation, one has to multiply an additional factor of $1/2$. The derived one-loop beta-coefficients using the particle content of the present framework are found to be $b_{2L} = -19/6$, $b_{Y} = 41/10$, $b_{3C} = -7$ from the SM to the LR breaking scale and $b_{2L} = b_{2R} = -13/6+2/3$, 
$b_{BL}= 59/6+3/2+3/2$, $b_{3C} = -17/3+0$ from the LR breaking scale to the GUT scale. The unification of gauge couplings is displayed in Fig.~\ref{fig:unifn} with the breaking scales
\begin{align}
	M_\text{GUT} = 10^{15.75}\text{ GeV}, \quad M_\text{R} = 8\text{ TeV}.
\end{align}
%

\section{Neutrinoless double beta decay}
\label{secbeta}
As discussed earlier, there is no tree level Dirac neutrino mass term connecting light and heavy neutrinos. Consequently, the mixing between light and heavy neutrinos is vanishing at this order. Also, the mixing between the charged gauge bosons is vanishing at the tree level due to the absence of a scalar bidoublet. 

The charged current interaction in the mass basis for the leptons is given by 
\begin{align}
 \frac{g_L}{\sqrt{2}} \sum_{i=1}^3 U_{ei} \left(
 \overline{\ell_{L}} \gamma_\mu \nu_i W^{\mu}_L + \frac{g_R}{g_L} 
 \overline{\ell}_{R} \gamma_\mu N_i   W^{\mu}_R \right) + \text{h.c.} 
\end{align}
The charged current interaction for leptons leads to $0\nu\beta\beta$ decay via the exchange of light and heavy neutrinos. There are additional contributions to $0\nu\beta\beta$ decay due to doubly charged triplet scalar exchange. While the left-handed triplet exchange is suppressed because of its small induced VEV, the right-handed triplet can contribute sizeably to $0\nu\beta\beta$ decay. 

Before numerical estimation, let us point out the mass relations between light and heavy neutrinos under natural type-II seesaw dominance. For a  hierarchical pattern of light neutrinos the mass eigenvalues are given as $ m_1 < m_2 \ll m_3$. The lightest neutrino mass eigenvalue is $m_1$ while the other mass eigenvalues are determined using the oscillation parameters as follows, $m^2_2 = m_1^2 +\Delta m_{\rm sol}^2$, $m^2_3 = m_1^2 +\Delta m_{\rm atm}^2 + \Delta m_{\rm sol}^2$.
On the other hand, for the inverted hierarchical pattern of the light neutrino masses $m_3 \ll m_1 \approx m_2$ where $m_3$ is the lightest mass eigenvalue while other mass eigenvalues are determined by $m^2_1 = m_3^2 +\Delta m_{\rm atm}^2$, $m^2_2 = m_3^2 +\Delta m_{\rm sol}^2 +\Delta m_{\rm atm}^2$. The quasi-degenerate pattern of light neutrinos is $m_1 \approx m_2 \approx m_3 \gg \sqrt{\Delta m^2_\text{atm}}$. In any case, the heavy neutrino masses are directly proportional to the light neutrino masses.

In the present analysis, we discuss $0\nu\beta\beta$ decay due to exchange of light neutrinos via left-handed currents, right-handed neutrinos via right-handed currents as well as a right-handed doubly charged scalar\footnote{A detailed discussion of $0\nu\beta\beta$ decay within LRSMs can be found e.g. in Refs.~\cite{Mohapatra:1980yp, Mohapatra:1981pm, Hirsch:1996qw, Tello:2010am, Chakrabortty:2012mh, Patra:2012ur, Awasthi:2013ff, Barry:2013xxa, Dev:2013vxa, Ge:2015yqa, Awasthi:2015ota}.}. The half-life for a given isotope for these contributions is given by
\begin{align}
 [T_{1/2}^{0\nu}]^{-1} \!=\! G_{01} 
 \left(|\mathcal{M}_\nu \eta_{\nu}|^2 + |\mathcal{M}'_N \eta_N + \mathcal{M}_N \eta_\Delta)|^2\right),
\end{align}
where $G_{01}$ corresponds to the standard $0\nu\beta\beta$ phase space factor, the $\mathcal{M}_i$ correspond to the nuclear matrix elements for the different exchange processes and $\eta_i$ are dimensionless parameters determined below. 

\subparagraph*{Light neutrinos} The lepton number violating dimensionless particle physics parameter derived from $0\nu\beta\beta$ decay due to the standard mechanism via the exchange of light neutrinos is
\begin{align}
\label{eta:nu} 
\mathcal{\eta}_{\nu} =\frac{1}{m_e}  \sum^{3}_{i=1} U^2_{ei}\, m_{i}
           = \frac{m^\nu_{\rm ee}}{m_e} \,. 
\end{align}
Here, $m_e$ is the electron mass and the effective $0\nu\beta\beta$ mass is explicitly given by
\begin{align}
\label{eq:mee-std}
m^\nu_{\rm ee}
=\left| c^2_{12} c^2_{13} m_1 + s^2_{12} c^2_{13} m_2 e^{i\alpha} + s^2_{13} m_3 e^{i\beta} \right| \,,
\end{align}
with the sine and cosine of the oscillation angles $\theta_{12}$ and $\theta_{13}$, $c_{12} = \cos\theta_{12}$, etc. and the unconstrained Majorana phases $0 \leq \alpha,\beta < 2\pi$.

\subparagraph*{Right-handed neutrinos}
The contribution to $0\nu\beta\beta$ decay arising from the purely right-handed currents via the exchange of right-handed neutrinos generally results in the lepton number violating dimensionless particle physics parameter
\begin{align}
\label{eta:N} 
\mathcal{\eta}_N &= 
	m_p \left(\frac{g_R}{g_L}\right)^4 \left(\frac{M_{W_L}}{M_{W_R}}\right)^4  
              \sum^{3}_{i=1} \frac{U^2_{ei} M_i}{|p|^2 + M^2_i}\,.
\end{align}
The virtual neutrino momentum $|p|$ is of the order of the nuclear Fermi scale, $p\approx 100$~MeV. $m_p$ is the proton mass and for the manifest LRSM case we have $g_L = g_R$, or else the new contributions are rescaled by the ratio between these two couplings. We in general consider right-handed neutrinos that can be either heavy or light compared to nuclear Fermi scale.  

If the mass of the exchanged neutrino is much higher than its momentum, $M_i \gg |p|$, the propagator simplifies as 
\begin{align}
\frac{M_i}{p^2-M_i^2} \approx -\frac{1}{M_i},
\end{align}
and the effective parameter for right-handed neutrino exchange yields
\begin{align}
\eta_N =m_p\left(\frac{g_R}{g_L}\right)^4\left(\frac{M_W}{M_{W_R}}\right)^4
\sum_{i=1}^3 \frac{U^2_{ei}}{M_i} \propto \eta_{\nu}(m_i^{-1})\,,
\end{align} 
where in the expression for $\eta_\nu(m_i^{-1})$ the individual neutrino masses are replaced by their inverse values. Such a contribution clearly becomes suppressed the smaller the right-handed neutrino masses are.

On the other hand, if the mass of the neutrino is much less than its typical momentum, 
$M_i \ll |p|$, the propagator simplifies in the same way as for the light neutrino exchange,
\begin{align}
   P_{R}\frac{\slashed{p}+M_i}{p^2-M_i^2}P_{R} \approx \frac{M_i}{p^2}\,,
\end{align}
because both currents are right-handed. As a result, the $0\nu\beta\beta$ decay contribution leads to the dimensionless parameter
\begin{align}
	\eta_N =\frac{m_p}{|p|^2}\left(\frac{g_R}{g_L}\right)^4\left(\frac{M_W}{M_{W_R}}\right)^4
             \sum_{i=1}^3 U^2_{ei} M_i \propto \eta_\nu \,.
\end{align} 
This is proportional to the standard parameter $\eta_\nu$ but in the case of very light right-handed neutrinos, e.g. $M_i \approx m_i$, the contribution becomes negligible because of the strong suppression with the heavy right-handed $W$ boson mass.

In general, we consider right-handed neutrinos both lighter and heavier than 100~MeV and use \eqref{eta:N} to calculate the contribution. In addition, the relevant nuclear matrix element changes; for $M_i \gg 100$~MeV it approaches $\mathcal{M}'_N \to \mathcal{M}_N$ whereas for $M_i \ll 100$~MeV it approaches $\mathcal{M}'_N \to \mathcal{M}_\nu$. For intermediate values, we use a simple smooth interpolation scheme within the regime 10~MeV -- 1~GeV, which yields a sufficient accuracy for our purposes.

\subparagraph*{Right-handed triplet scalar}
Finally, the exchange of a doubly charged right-handed triplet scalar gives
\begin{align}
\label{eta:delta} 
\mathcal{\eta}_\Delta = \frac{m_p}{M^2_{\delta_R^{--}}} \left(\frac{g_R}{g_L}\right)^4 \left(\frac{M_W}{M_{W_R}}\right)^4 \sum^{3}_{i=1} U^2_{ei} M_i \propto \eta_\nu\,.
\end{align}
This expression is also proportional to the standard $\eta_\nu$ because the relevant coupling of the triplet scalar is proportional to the right-handed neutrino mass.

\begin{figure}[t!]
	\includegraphics[width=0.45\textwidth]{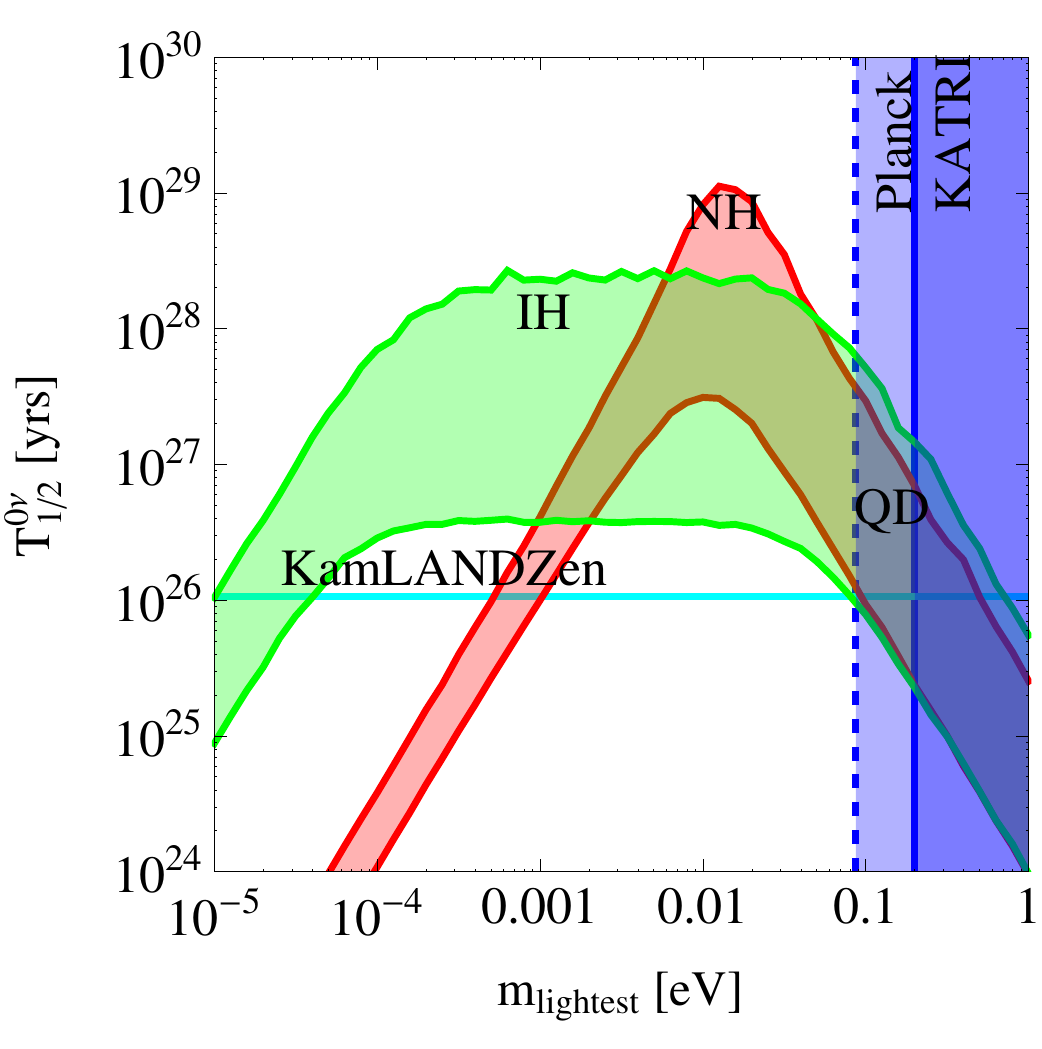}
	\caption{$0\nu\beta\beta$ decay half-life as a function of the lightest neutrino mass in the case of normal hierarchical (NH) and inverse hierarchical (IH) light neutrinos in red and green bands respectively. The other parameters are fixed as $M_{W_R} = 5$~TeV, $M_{\delta^{--}_R} \approx 5$~TeV and the heaviest right-handed neutrino mass is 1~TeV. The gauge couplings are assumed universal, $g_L=g_R$, and the intermediate values for the nuclear matrix elements are used, $\mathcal{M}_\nu = 4.5$, $\mathcal{M}_N = 270$. The bound on the sum of light neutrino masses from the KATRIN and Planck experiments are represented as vertical lines. The bound from KamLAND-Zen experiment is presented in horizontal line for Xenon isotope. The bands arise due to $3\sigma$ range of neutrino oscillation parameters and variation in the Majorana phases from $0-2\pi$.} 
	\label{plot:0nubb-Total}
\end{figure}

\subparagraph*{Numerical estimate}
\begin{table}[t!]
	\centering
	\begin{tabular}{lccc}
		\hline
		Isotope & $G_{01}~({\rm yr}^{-1})$ & ${\cal M}_\nu$ & ${\cal M}_N$  \\
		\hline 
		$^{\phantom{1}76}$Ge   & $5.77 \times 10^{-15}$ & 2.58--6.64 & 233--412  \\ 
		$^{136}$Xe  & $3.56 \times 10^{-14}$ & 1.57--3.85 & 164--172  \\
		\hline
	\end{tabular}
	\caption{Phase space factor $G_{01}$ and ranges of nuclear matrix elements for light and heavy neutrino exchange for the isotopes $^{76}$Ge and $^{136}$Xe \cite{Meroni:2012qf}.}
	\label{tab:nucl-matrix}
\end{table}
In the following, we numerically estimate the half-life for $0\nu\beta\beta$ decay of the isotope
$^{136}$Xe as shown in Fig.\ref{plot:0nubb-Total}. We use the current values of masses and mixing parameters 
from neutrino oscillation data reported in the global fits taken from Ref.~\cite{GonzalezGarcia:2012sz}. 
For the $0\nu\beta\beta$ phase space factors and nuclear matrix elements we use the values given 
in Tab.~\ref{tab:nucl-matrix}\,. In Fig.~\ref{plot:0nubb-Total}, we show the dependence of the $0\nu\beta\beta$ decay half-life 
on the lightest neutrino mass, i.e. $m_1$ for normal and $m_3$ for inverse hierarchical 
neutrinos. The other model parameters are fixed as 
\begin{gather}
g_R = g_L, M_{W_R}=M_{\delta_R^{--}} \approx 5\text{ TeV}\,,
M_N^\text{heaviest} = 1\text{ TeV} \,.
\end{gather}
The lower limit on lightest neutrino mass is derived to be $m_< \approx$ 0.9 meV, 0.01 meV 
for NH and IH pattern of light neutrino masses respectively by saturating the KamLAND-Zen 
experimental bound.

As for the experimental constraints, we use the current best limits at 90\% C.L., $T_{1/2}^{0\nu}(^{136}\text{Xe}) > 1.07\times 10^{26}$~yr and $T_{1/2}^{0\nu}(^{76}\text{Ge}) > 2.1\times 10^{25}$~yr from KamLAND-Zen~\cite{KamLAND-Zen:2016pfg} and the GERDA Phase I~\cite{Agostini:2013mzu}, respectively. Representative for the sensitivity of future $0\nu\beta\beta$ experiments, we use the expected reach of the planned nEXO experiment, $T_{1/2}^{0\nu}(^{136}\text{Xe}) \approx 6.6\times 10^{27}$~yr \cite{Albert:2014afa}. As for the other experimental probes on the neutrino mass scale, we use the future sensitivity of the KATRIN experiment on the effective single $\beta$ decay mass $m_\beta \approx 0.2$~eV \cite{Mertens:2015ila} and the current limit on the sum of neutrino masses from cosmological observations, $\Sigma_i m_i \lesssim 0.7$~eV~\cite{Ade:2015xua}.

\begin{figure*}[t!]
\includegraphics[width=0.48\textwidth]{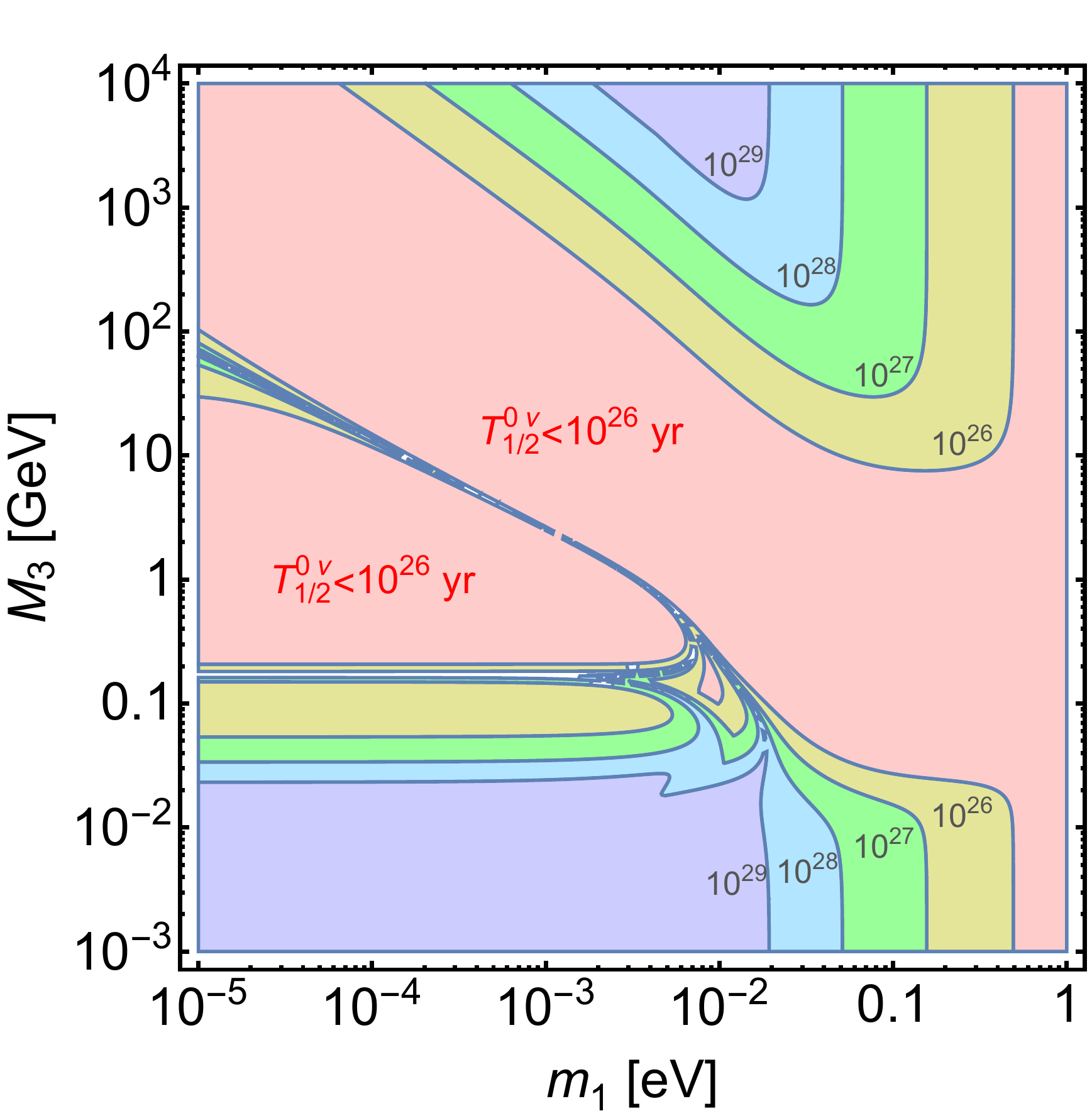}
\includegraphics[width=0.48\textwidth]{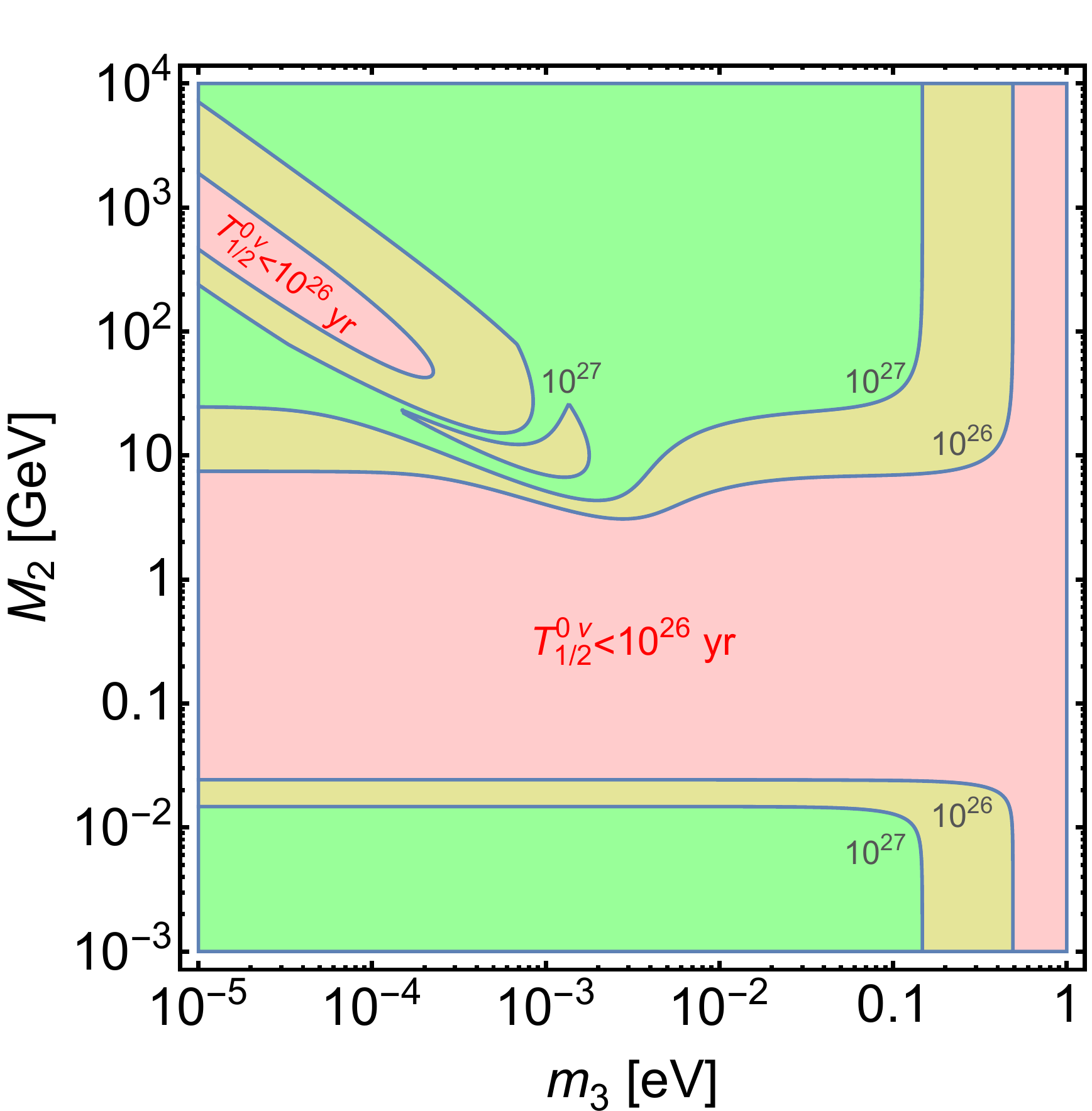}
\caption{Half-life of $0\nu\beta\beta$ decay in Xe as a function of the lightest and the heaviest neutrino mass for a normal (left) and inverse (right) neutrino mass hierarchy. The contours denote the half-life in years. Best-fit oscillation data are used and the Majorana phases are chosen to yield the longest half-life. Likewise, the smallest values of the nuclear matrix elements in Tab.~\ref{tab:nucl-matrix} are employed. The other model parameters are chosen as $g_R = g_L$ and $M_{W_R} = M_\Delta = 5$~TeV.} 
\label{plot:0nubb-contour}
\end{figure*}
For a better understand of the interplay between the left- and right-handed neutrino mass scales, we show in Fig.~\ref{plot:0nubb-contour} the $0\nu\beta\beta$ decay half-life as a function of the lightest neutrino mass and the heaviest neutrino mass for a normal (left) and inverse (right) neutrino mass hierarchy. The other model parameters are fixed, with right-handed gauge boson and doubly-charged scalar masses of 5~TeV. The oscillation parameters are at their best fit values and the Majorana phases are always chosen to yield the smallest rate at a given point, i.e. the longest half life. The nuclear matrix employed are at the lower end in Tab.~\ref{tab:nucl-matrix}. This altogether yields the longest, i.e. most pessimistic, prediction for the $0\nu\beta\beta$ decay half-life. The red-shaded area is already excluded with a predicted half life of $10^{26}$~yr or faster. As expected, this sets an upper limit on the lightest neutrino mass $m_\text{lightest} \lesssim 1$~eV, but it also puts stringent constraints on the mass scale of the right-handed neutrinos. For an inverse hierarchy, the range $50\text{ MeV} \lesssim M_2 \lesssim 5$~GeV is excluded whereas in the normal hierarchy case, large $M_3$ can be excluded if there is a strong hierarchy, $m_1 \to 0$. This is due the large contribution of the lightest heavy neutrino $N_1$ in such a case.

\section{Conclusion}
\label{conclusion}
We have presented a Left-Right symmetric model with additional vector-like fermions in order to simultaneously explain the charged fermion and Majorana neutrino masses. The quark and charged lepton masses and mixing is realized via a universal seesaw mechanism. Although spontaneous symmetry breaking is achieved with two doublet Higgs fields with non-zero $B-L$ charge, we have introduced scalar triplets with small induced VEVs such that they give Majorana masses to light as well as heavy neutrinos. The Majorana nature of these neutrinos leads to $0\nu\beta\beta$ decay and it is found that the right-handed currents play an important role in discriminating between the mass hierarchy as well as the absolute scale of light neutrinos. We have also embedded the framework in a non-supersymmetric $SO(10)$ GUT and and found that the gauge couplings unify at a scale $10^{15.75}$~GeV when we introduce a scalar bitriplet at the left-right breaking scale. 

\section*{Acknowledgements}
PP is supported by the DST INSPIRE Fellowship (No. DST/INSPIRE Fellowship/2014/IF140299) funded by the Department 
of Science and Technology, India. The work of SP is partially supported by the Department of Science and Technology, 
Govt.\ of India under the financial grant SB/S2/HEP-011/2013. The work of US is supported partly by the JC Bose National Fellowship grant under DST, India.

\bibliography{universal_lrsm}
\bibliographystyle{utcaps_mod}
\end{document}